\documentclass[preprint,preprintnumbers, prd, floatfix,  superscriptaddress,nofootinbib] {revtex4-1}
\usepackage{graphicx}
\usepackage{epsfig}
\usepackage{bm}
\usepackage{amssymb}
\usepackage{float}
\usepackage{amsmath}
\usepackage{dcolumn}
\usepackage{cancel}
\usepackage[colorlinks]{hyperref}
\usepackage[usenames,dvipsnames]{color}
\usepackage{color}
\usepackage{epstopdf}
\usepackage{nccbbb}
\usepackage{setspace}

\newcommand{\CD}{{\cal D}}
\newcommand{\CR}{{\cal R}}
\newcommand{\CQ}{{\cal Q}}

\newcommand{\average}[1]{\left\langle #1 \right\rangle_\CD}

\newcommand{\naverage}[1]{\left\langle #1 \right\rangle_{\CD_{\rm \bf 0}}}

\newcommand{\now}[1]{{#1_{\rm \bf 0}}}
\newcommand{\be}{\begin{equation}}
\newcommand{\ee}{\end{equation}}
\newcommand{\bea}{\begin{eqnarray}}
\newcommand{\eea}{\end{eqnarray}}
\newcommand{\bean}{\begin{eqnarray*}}
\newcommand{\eean}{\end{eqnarray*}}
\hypersetup{
     breaklinks=true,
    pdfstartview={FitH},    colorlinks=true,       % false: boxed links; true: colored links
    linkcolor=blue,          % color of internal links
    citecolor=red,        % color of links to bibliography
    filecolor=magenta,      % color of file links
    urlcolor=blue,           % color of external links
    anchorcolor=green,      % Color for anchor text
    linktocpage=true
}
\begin{document}
\title{Testing backreaction effects with type Ia supernova data and observational Hubble parameter
data}

\author{Yan-Hong Yao}
\email{yhy@mail.nankai.edu.cn}
\author{Xin-He Meng}
\email{xhm@nankai.edu.cn}

\affiliation{Schcool of Physics, Nankai University, Tianjin 300071, China}

\begin{abstract}
The backreaction term ${\cal Q}_\CD$ and the averaged spatial Ricci scalar $\average{\CR}$ in the spatially averaged inhomogeneous Universe can be used to combine into effective perfect fluid energy density $\varrho_{\rm eff}^{\CD}$ and pressure $p_{\rm eff}^{\CD}$ that can be regarded as new effective sources for the backreaction effects. In order to model the realistic evolution of backreaction, we adopt the Chevallier-Polarski-Linder(CPL) parameterizations of the equation of state(EoS) of the effective perfect fluid. To deal with observations in the
backreaction context, in this paper, we employ two metrics to describe the the late time Universe, one of them is the standard Friedmann-Lema\^{\i}tre-Robertson-Walker(FLRW) metric, and the other is a template metric with an evolving curvature parameter introduced by Larena et. al. in \cite{larena2009testing}. We also fit the CPL backreaction model using type Ia supernova(SN Ia) data and observational Hubble parameter data(OHD) with these two metrics, and find out that parameter tensions between two different data sets are larger when the backreaction model is equipped with the template metric, therefore we conclude that the prescription of the geometrical instantaneous spatially-constant curvature $\kappa_{\CD}$ needs to be modified.

\end{abstract}

\maketitle

\section{Introduction}
\label{intro}
A large number of observations \cite{riess1998,perlmutter1999measurements,spergel2003first,spergel2007three,dunkley2009five,larson2011seven,hinshaw2013nine}, from low redshift to high redshift, suggest that the Universe is in a state of accelerated expansion, which implies that there exists a sector named as dark energy with negative pressure accounting for the invisible fuel that accelerates the expansion rate of the current universe. There are many scenarios have been proposed to account for these observations, the simplest one is the cosmological constant scenario, in which cosmological constant or, equivalently, vacuum energy plays the role of dark energy. However, because of the huge discrepancy between the theoretical expected vacuum energy density and the observed one, other alternative scenarios have been proposed, including scalar field models such as quintessence\cite{Caldwell1998Cosmological}, phantom\cite{Caldwell1999A}, dilatonic\cite{Piazza2004Dilatonic}, tachyon\cite{Padmanabhan2002Accelerated}and quintom\cite{Bo2006Oscillating} etc. and modified gravity models such as braneworlds \cite{maartens2010brane}, scalar-tensor gravity \cite{esposito2001scalar}, higher-order gravitational theories \cite{capozziello2005reconciling,das2006curvature}. Recently, a third alternative has been considered to explain the dark energy
phenomenon as backreaction effect(a Large scale effect caused by small scale inhomogeneities of the universe)\cite{rasanen2004dark,kolb2006cosmic}.

In order to consider backreaction effect, it is necessary to answer a longstanding question that how to average a general inhomogeneous spacetime.
To date the macroscopic gravity(MG) approach \cite{zalaletdinov1992averaging,zalaletdinov1993towards,mars1997space} is probably the most well known attempt at averaging in space-time. Although it provides a prescription for the correlation functions which emerge in an averaging of the Einstein's field equations, so far it required a number of assumptions about the correlation functions which make the theory less convictive. Therefore, many researchers adopt another averaged approach put forward by Buchert \cite{buchert2000average,buchert2001average}, in despite of its foliation dependent nature, such approach is quite simple and hence becomes the most well studied averaged approach. Since the averaged field equations in such approach do not form a closed set, one needs to make some assumptions about the backreaction term appeared in the averaged equations. In \cite{buchert2006correspondence}, by taking the assumption that the backreaction term ${\cal Q}_\CD$ and the averaged spatial Ricci scalar $\average{\CR}$ obey the scaling laws of the volume scale factor $a_\CD$, Buchert proposed a simple backreaction model(we will refer it as scaling backreaction model in the follow). In order to test the ability of such backreaction model to correctly describe observations of the large scale properties of the Universe, Larena et. al. in \cite{larena2009testing} introduced a template metric with an evolving curvature parameter and
confronted the scaling backreaction model with such template metric against supernova data and the position of the Cosmic Microwave Backround (CMB) peaks. Because of doubting the validity of the prescription for such evolving curvature parameter, Cao et. al.\cite{cao2018testing} fitted the scaling backreaction model using OHD with the FLRW metric and the template metric, and found out that, for the template metric case, their fitting results is in contrary with the constraint results of Larena et al.\cite{larena2009testing}, and for the FLRW metric case, two constraint results are basically consistent, which, to a certain extent, supporting their guess that the prescription for the geometrical instantaneous spatially-constant curvature $\kappa_{\CD}$ should be modified. However, as was pointed out by Larena et. al. in \cite{larena2009testing}, the pure scaling ansatz for ${\cal Q}_\CD$ and $\average{\CR}$ is not what we expected in a realistic evolution of backreaction, therefore the conclusion that the  prescription for $\kappa_{\CD}$ needs to be modified was drawn too early. In order to prove Cao et. al.'s conjecture in a more persuasive way, we describe the backreaction effects with an effective perfect fluid, whose EoS was assumed as $w_{eff}^{\CD}=w_{0}^{\CD}+w_{a}^{\CD}(1-a_\CD)$ which having one more parameter than that of the scaling backreaction model. Then we fit such CPL backreaction model using SN Ia data and OHD with the FLRW metric and the template metric, and the constraint results agree with Cao et. al.'s conjecture.

The paper is organized as follows. In Section \ref{sec:1}, the spatial averaged approach of Buchert is demonstrated with presentation of the averaged equations for the volume scale factor $a_\CD$, and under this theoretical framework, the CPL backreaction model is introduced. In Section \ref{sec:2}, we introduce the template metric proposed by Larena et. al., which is a necessary tool to test the theoretical predictions with observations, although we doubt the validity of the prescription for $\kappa_{\CD}$. We also show the way to compute observables with such template metric in this section. In Section \ref{sec:3}, we apply a likelihood analysis of the CPL backreaction model by confronting it using latest SN Ia data and OHD with the FLRW metric and the template metric. After analysis of the results in Section \ref{sec:3}, we summarize our results in the last section.
\section{The CPL backreaction model}
\label{sec:1}
In \cite{buchert2000average}, Buchert considers a universe filled with irrotational dust with energy density $\varrho$. By foliating space-time with the use of Arnowitt-Deser-Misner(ADM) procedure and defining an averaging operator that acts on any spatial scalar $\Psi$ function as
\begin{equation}
\average{\Psi}:=\frac{1}{V_{\CD}}\int_{\CD}\Psi Jd^{3}X,
\end{equation}
where $V_{\CD}:=\int_{\CD}Jd^{3}X$ is the domain's volume, Buchert obtains two averaged equations here we need, the averaged Raychaudhuri equation
\begin{equation}
3\frac{{\ddot a}_\CD}{a_\CD} + 4\pi G \average{\varrho} ={\CQ}_\CD,
\end{equation}
and the averaged Hamiltonian constraint
\begin{equation}
3\left( \frac{{\dot a}_\CD}{a_\CD}\right)^2 - 8\pi G \average{\varrho}= - \frac{\average{\CR}+{\CQ}_\CD }{2}.
\end{equation}
In these two equations, $a_\CD(t) = \left(\frac{V_\CD (t)}{V_{\now\CD}}\right)^{1/3}$ is the volume scale factor, where $V_{\now\CD} =\vert{\now\CD}\vert$ denotes the present value of the volume, and ${\cal Q}_\CD$, $\average{\CR}$  represent the backreaction term and the averaged spatial Ricci scalar respectively, which are related by the following integrability condition
\begin{equation}
\frac{1}{a_\CD^6}\partial_t \left(\,{\CQ}_\CD \,a_\CD^6 \,\right)
\;+\; \frac{1}{a_\CD^{2}} \;\partial_t \left(\,\average{\CR}a_\CD^2 \,
\right)\,=0\;.
\end{equation}
Now we can define the effective prefect fluid by
\begin{eqnarray}
  \varrho_{\rm eff}^{\CD}:&=& -\frac{1}{16\pi G}({\CQ}_\CD+\average{\CR}), \\
  p_{\rm eff}^{\CD}:&=&  -\frac{1}{16\pi G}({\CQ}_\CD-\frac{\average{\CR}}{3}),
\end{eqnarray}
 the averaged Raychaudhuri equation and the averaged Hamiltonian constraint then can formally be recast into standard Friedmann equations for a total perfect fluid energy momentum tensor
 \begin{equation}\label{7}
   3\frac{{\ddot a}_\CD}{a_\CD} + 4\pi G (\average{\varrho}+\varrho_{\rm eff}^{\CD}+3p_{\rm eff}^{\CD} )= 0,
 \end{equation}
 \begin{equation}\label{8}
    3\left( \frac{{\dot a}_\CD}{a_\CD}\right)^2 = 8\pi G (\average{\varrho}+\varrho_{\rm eff}^{\CD}),
 \end{equation}
given the effective energy density and pressure, the effective EoS reads
\begin{equation}\label{}
w_{\rm eff}^{\CD}:=\frac{p_{\rm eff}^{\CD}}{\varrho_{\rm eff}^{\CD}}=\frac{{\CQ}_\CD-\frac{\average{\CR}}{3}}{{\CQ}_\CD+\average{\CR}}.
\end{equation}
One can then obtain a specific backreaction model with an extra ansatz about the form of $w_{\rm eff}^{\CD}$, for example, for the scaling backreaction model, we have $w_{\rm eff}^{\CD}=-\frac{n+3}{3}$, in which $n$ is the scaling index. In this paper, in order to model the realistic evolution of backreaction, we assume that the effective EoS follows the CPL form, i.e.
\begin{equation}\label{}
  w_{\rm eff}^{\CD}=w_{0}^{\CD}+w_{a}^{\CD}(1-a_\CD),
\end{equation}
so the Eq.~(\ref{8}) can be rewrite as
\begin{equation}\label{}
  H_{\CD}^{2}=H_{\now\CD}^{2}[\Omega_{m}^{\now\CD}a_{\CD}^{-3}+\Omega_{X}^{\now\CD}a_{\CD}^{-3(1+w_{0}^{\CD}+w_{a}^{\CD})}e^{-3w_{a}^{\CD}(1-a_{\CD})}],
\end{equation}
here $H_\CD : = {\dot a}_\CD / a_\CD$ denotes the volume Hubble parameter and $\Omega_{m}^{\now\CD}:= \frac{8\pi G}{3 H_{\now\CD}^2} \naverage{\varrho}$, $\Omega_{X}^{\now\CD}:= \frac{8\pi G}{3 H_{\now\CD}^2}\varrho_{\rm eff}^{\now\CD}=1-\Omega_{m}^{\now\CD}$.
Combine the above equations, we have the following formulas for the backreaction term and the averaged spatial Ricci scalar
\begin{equation}\label{}
{\cal Q}_\CD=-\frac{9}{2}(1-\Omega_{m}^{\now\CD})H_{\now\CD}^{2}e^{-3w_{a}^{\CD}(1-a_{\CD})}[(\frac{1}{3}+w_{0}^{\CD}+w_{a}^{\CD})a_{\CD}^{-3(1+w_{0}^{\CD}+w_{a}^{\CD})}-w_{a}^{\CD}a_{\CD}^{-2-3(w_{0}^{\CD}+w_{a}^{\CD})}],
\end{equation}
\begin{equation}\label{}
  \average{\CR}=-\frac{9}{2}(1-\Omega_{m}^{\now\CD})H_{\now\CD}^{2}e^{-3w_{a}^{\CD}(1-a_{\CD})}[(1-w_{0}^{\CD}-w_{a}^{\CD})a_{\CD}^{-3(1+w_{0}^{\CD}+w_{a}^{\CD})}+w_{a}^{\CD}a_{\CD}^{-2-3(w_{0}^{\CD}+w_{a}^{\CD})}].
\end{equation}

\section{Effective geometry}
\label{sec:2}
\subsection{The template metric}
In order to test the ability of spatially averaged inhomogeneous cosmology to correctly describe observations of the large scale properties of the Universe, Larena et. al. in \cite{larena2009testing} introduced a template metric as follows,
\begin{equation}
\label{eq:tempmetric1}
{}^4 {\bf g}^\CD = -dt^2 + L_{\now H}^2 \,a_\CD^2 \gamma^\CD_{ij}\,dX^i \otimes dX^j \;\;,
\end{equation}
where $L_{\now H}=1/H_{\now\CD}$ is the present size of the horizon introduced so that the coordinate distance is dimensionless,
and the domain-dependent effective three-metric reads:
\begin{equation}
\gamma^\CD_{ij}\,dX^i \otimes dX^j =\frac{dr^2}{1-\kappa_{\CD}(t)r^2}+r^2d\Omega^{2},
\end{equation}
with $d\Omega^{2}=d\theta^{2}+\sin^{2}(\theta)d\phi^2$, this effective three-template metric is identical to the spatial part of a FLRW metric at any given time, but its scalar curvature  $\kappa_{\CD}$  can vary from time to time. As was pointed out by Larena et. al., $\kappa_{\CD}$ cannot be arbitrary, more precisely, they argue that it should be related to the true averaged scalar curvature $\average{\CR}$ in the way that
\begin{equation}
\average{\CR}=\frac{\kappa_{\CD}(t)|\naverage{\CR}|a_{\now\CD}^{2}}{a_{\CD}^{2}(t)}.
\end{equation}
However, considering the fitting results from Cao et. al. \cite{cao2018testing}, we suspect that this prescription is incorrect, this is the reason why we start this research.

In our CPL backreaction model, we have the formula for $\kappa_{D}$ as follows,
\begin{equation}\label{}
\kappa_{\CD}(t)=-\frac{\Omega_{X}^{\now\CD}}{|\Omega_{X}^{\now\CD}(1-w_{0}^{\CD})|}e^{-3w_{a}^{\CD}(1-a_{\CD})}[(1-w_{0}^{\CD}-w_{a}^{\CD})a_{\CD}^{-1-3(w_{0}^{\CD}+w_{a}^{\CD})}+w_{a}^{\CD}a_{\CD}^{-3(w_{0}^{\CD}+w_{a}^{\CD})}].
\end{equation}

\subsection{Computation of observables}
The computation of effective distances along the light cone defined by the template metric is very different from that of distances in FLRW models. Firstly, let us introduce an effective redshift $z_{\CD}$ defined by
\begin{equation}
1+z_{\CD}:=\frac{(g_{ab}k^{a}u^{b})_{S}}{(g_{ab}k^{a}u^{b})_{O}}\mbox{ ,}
\end{equation}
where the letters O and S denote the evaluation of the quantities at the observer and at the source respectively, $g_{ab}$ in this expression represents the template metric, while $u^{a}$ is the four-velocity of the dust which satisfies $u^{a}u_{a}=-1$, $k^{a}$ the wave vector of a light ray travelling from the source S towards the observer O with the restrictions $k^{a}k_{a}=0$. Then, by normalizing this wave vector such that $(k^{a}u_{a})_{O}=-1$ and introducing the scaled vector $\hat{k}^{a}=a_{\CD}^{2}k^{a}$, we have the following equation:
\begin{equation}
\label{eq:defred2}
1+z_{\CD}=(a_{\CD}^{-1}\hat{k}^{0})_{S}\mbox{ ,}
\end{equation}
with $\hat{k}^{0}$ obeying the null geodesics equation $k^{a}\nabla_{a}k^{b}=0$ which leads to
\begin{equation}
\label{eq:evolk}
\frac{1}{\hat{k}^{0}}\frac{d\hat{k}^{0}}{da_{\CD}}=-\frac{r^{2}(a_{\CD})}{2(1-\kappa_{\CD}(a_{\CD})r^{2}(a_{\CD}))}\frac{d\kappa_{\CD}(a_{\CD})}{da_{\CD}}\mbox{ .}
\end{equation}

As usual, the coordinate distance can be derived from the equation of radial null geodesics:
\begin{equation}
\label{eq:coorddist}
\frac{dr}{da_{\CD}}=-\frac{H_{\now\CD}}{a_{\CD}^{2}H_{\CD}(a_{\CD})}\sqrt{1-\kappa_{\CD}(a_{\CD})r^{2}}
\end{equation}

Solving these two equations with the initial condition $\hat{k}^{0}(1)=1, r(1)=0 $ and then plugging $\hat{k}^{0}(a_\CD)$ into Eq.~(\ref{eq:defred2}), one finds the relation between the redshift and the scale factor. With these results, we can determine the volume Hubble parameter $H_{\CD}(z_{\CD})$ and the luminosity distance $d_{L}(z_{\CD})$ of the sources defined by the following formula
\begin{eqnarray}
\label{eq:distances}
d_{L}(z_{\CD})&=&\frac{1}{H_{\now\CD}}(1+z_{\CD})^{2}a_{\CD}(z_{\CD})r(z_{\CD}).
\end{eqnarray}
Having computed these two observables , it is then possible to compare the CPL backreaction model predictions with type Ia supernova and Hubble parameter observations.

\section{Constraints from SN Ia data and OHD}
\label{sec:3}
In this section, we perform a likelihood analysis on the parameters of the CPL backreaction model with both the FLRW metric and the template metric using SN Ia data and OHD respectively.
\begin{center}
  1.SN Ia data
\end{center}
We use the Pantheon sample\cite{scolnic2018complete} with 1048 data points to fit the model, and its $\chi^2$ function is
\begin{equation}\label{}
 \tilde{\chi}_{SN}^{2}(\Omega_{m}^{\now\CD},w_0^{\CD},w_a^{\CD})= \sum_{i=1}^{1048}\frac{\left(\mu_{obs_i}-5\log_{10}\left[H_{\now\CD}d_{L}({z_{\CD}}_{i})\right]-\mu_0\right)^{2}}{\sigma_{\mu_i}^{2}},
\end{equation}
where $\mu_{obs_i}$ represents the observed distance modulus, $\sigma_{\mu_i}$ denotes its statistical uncertainty and $\mu_0$ is a free parameter.

We can rewrite the $\chi^2$ as
\begin{equation}\label{25}
\tilde{\chi}_{SN}^{2}(\Omega_{m}^{\now\CD},w_0^{\CD},w_a^{\CD})=R-2\mu_0S+\mu_0^2T,
\end{equation}
in which $R$, $S$ and $T$ is defined as
\begin{eqnarray}
  R &=& \sum_{i=1}^{1048}\frac{\left(\mu_{obs_i}-5\log_{10}\left[H_{\now\CD}d_{L}({z_{\CD}}_{i})\right]\right)^{2}}{\sigma_{\mu_i}^{2}},\\
  S &=& \sum_{i=1}^{1048}\frac{\left(\mu_{obs_i}-5\log_{10}\left[H_{\now\CD}d_{L}({z_{\CD}}_{i})\right]\right)}{\sigma_{\mu_i}^{2}}, \\
  T &=& \sum_{i=1}^{1048}\frac{1}{\sigma_{\mu_i}^{2}}.
\end{eqnarray}
Since $\mu_0$ is an irrelevant parameter, we can get rid of it by marginalizing it, or we can find the extreme point of the $\chi^2$ with respect to the independent variable $\mu_0$ and substitute it in Eq.~(\ref{25}). Here we choose the second method, in fact both methods give similar results. Since the extreme point of the $\chi^2$ with respect to the independent variable $\mu_0$ is $\mu_0=S/T$, we have the final formula for the
$\chi^2$ as follows,
\begin{equation}\label{}
   \chi_{SN}^{2}(\Omega_{m}^{\now\CD},w_0^{\CD},w_a^{\CD})= R-\frac{S^{2}}{T}.
\end{equation}

\begin{center}
  2.OHD
\end{center}
For the observed Hubble parameter dataset in Table \ref{tab:1}, the best-fit values of the parameters $(H_{\now\CD},\Omega_{m}^{\now\CD},w_{0}^{\CD},w_{a}^{\CD})$ can be determined by a likelihood analysis based on the calculation of
\begin{equation}
 \chi^{2}_{H}=\sum_{i=1}^{38} \frac{(H_{\now\CD}E(z_{\CD_i};\Omega_{m}^{\now\CD},w_0^{\CD},w_a^{\CD})-H_{i})^2}{\sigma_{i}^{2}}.
\end{equation}

Here we assume that each Hubble parameter data point is independent. However, the covariance matrix of data is not necessarily diagonal, as discussed in \cite{yu2013nonparametric}, and if it is not, the problem will become complicated and should be treated by means of the method mentioned by \cite{yu2013nonparametric}.

Since $\chi_{SN}^{2}$ does not contain $H_{\now\CD}$, in order to directly compare the two sets of fitting results, we can marginalize the $H_{\now\CD}$ in the $\chi^{2}_{H}$ by using the following formula:
\begin{equation}\label{}
\begin{split}
  P(\Omega_{m}^{\now\CD},w_0^{\CD},w_a^{\CD}|\{H_{i}\})&=\int P(\Omega_{m}^{\now\CD},w_0^{\CD},w_a^{\CD},H_{\now\CD}|\{H_{i}\})dH_{\now\CD}\\
  &=\int\mathcal{L}(\{H_{i}\}|\Omega_{m}^{\now\CD},w_0^{\CD},w_a^{\CD},H_{\now\CD})P(H_{\now\CD})dH_{\now\CD},
\end{split}
\end{equation}
where $\mathcal{L}(\{H_{i}\}|\Omega_{m}^{\now\CD},w_0^{\CD},w_a^{\CD},H_{\now\CD})=(\prod_{i=1}^{38}\frac{1}{\sqrt{2\pi\sigma_{i}^{2}}})\exp(-\frac{\chi^{2}_{H}}{2})$ is the likelihood function before $H_{\now\CD}$ is marginalized, $P(H_{\now\CD})=\frac{1}{\sqrt{2\pi\sigma_{H}^{2}}}\exp[-\frac{(H_{\now\CD}-\mu_{H})^2}{\sigma_{H}^{2}}]$ is the prior of $H_{\now\CD}$, where $\mu_{H}=73.8$ and $\sigma_{H}=1.1$ \cite{wong2020h0licow}. After marginalization, we can get the probability distribution of the three parameters $\Omega_{m}^{\now\CD}$, $w_0^{\CD}$ and $w_a^{\CD}$ as
\begin{equation}
P(\Omega_{m}^{\now\CD},w_0^{\CD},w_a^{\CD}|\{H_{i}\}) = \frac{1}{\sqrt{A}}[{\rm erf}(\frac{B}{\sqrt{A}})+1]e^{\frac{B^{2}}{A}},
\end{equation}
where
\begin{equation}
  A=\frac{1}{2\sigma_{H}^{2}}+\sum_{i=1}^{38}\frac{E^2(z_{{\CD}_i};\Omega_{m}^{\now\CD},w_0^{\CD},w_a^{\CD})}{2\sigma_{i}^{2}}, B=\frac{\mu_{H}}{2\sigma_{H}^{2}}+\sum_{i=1}^{38}\frac{E(z_{{\CD}_i};\Omega_{m}^{\now\CD},w_0^{\CD},w_a^{\CD})H_i}{2\sigma_{i}^{2}}.
\end{equation}
By which we can obtain the $\chi^2$ of OHD after $H_{\now\CD}$ is marginalized.
\begin{equation}\label{}
   \chi_{OHD}^{2}(\Omega_{m}^{\now\CD},w_0^{\CD},w_a^{\CD})=-2\ln(P(\Omega_{m}^{\now\CD},w_0^{\CD},w_a^{\CD}|\{H_{i}\}))
\end{equation}

\begin{figure}[h]
%\begin{tabular}{cc}
\begin{minipage}{0.45\linewidth}
  \centerline{\includegraphics[width=1\textwidth]{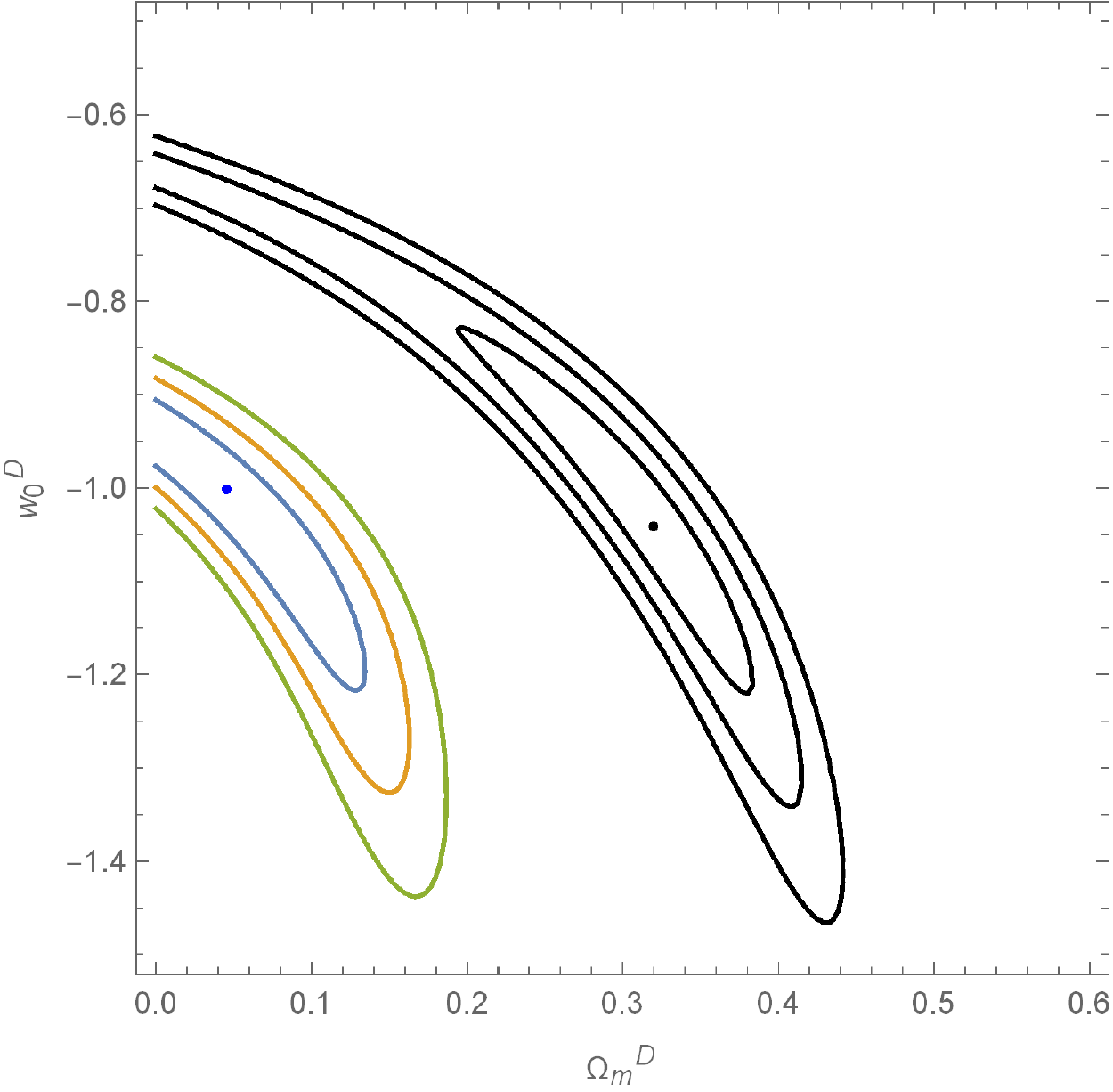}}
  \label{a}
\end{minipage}
\begin{minipage}{0.45\linewidth}
  \centerline{\includegraphics[width=1\textwidth]{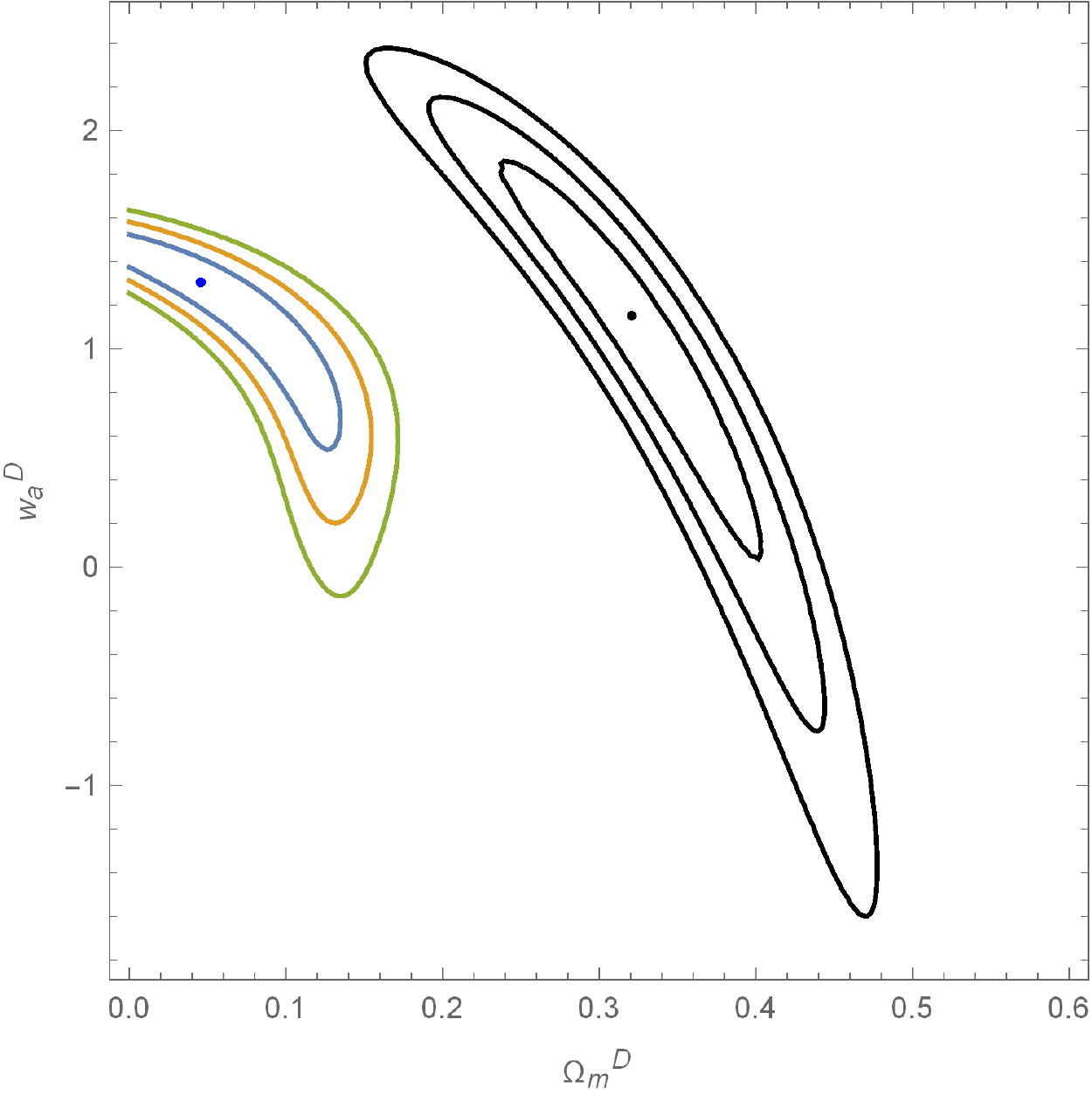}}
  \label{a}
\end{minipage}
\begin{minipage}{0.45\linewidth}
  \centerline{\includegraphics[width=1\textwidth]{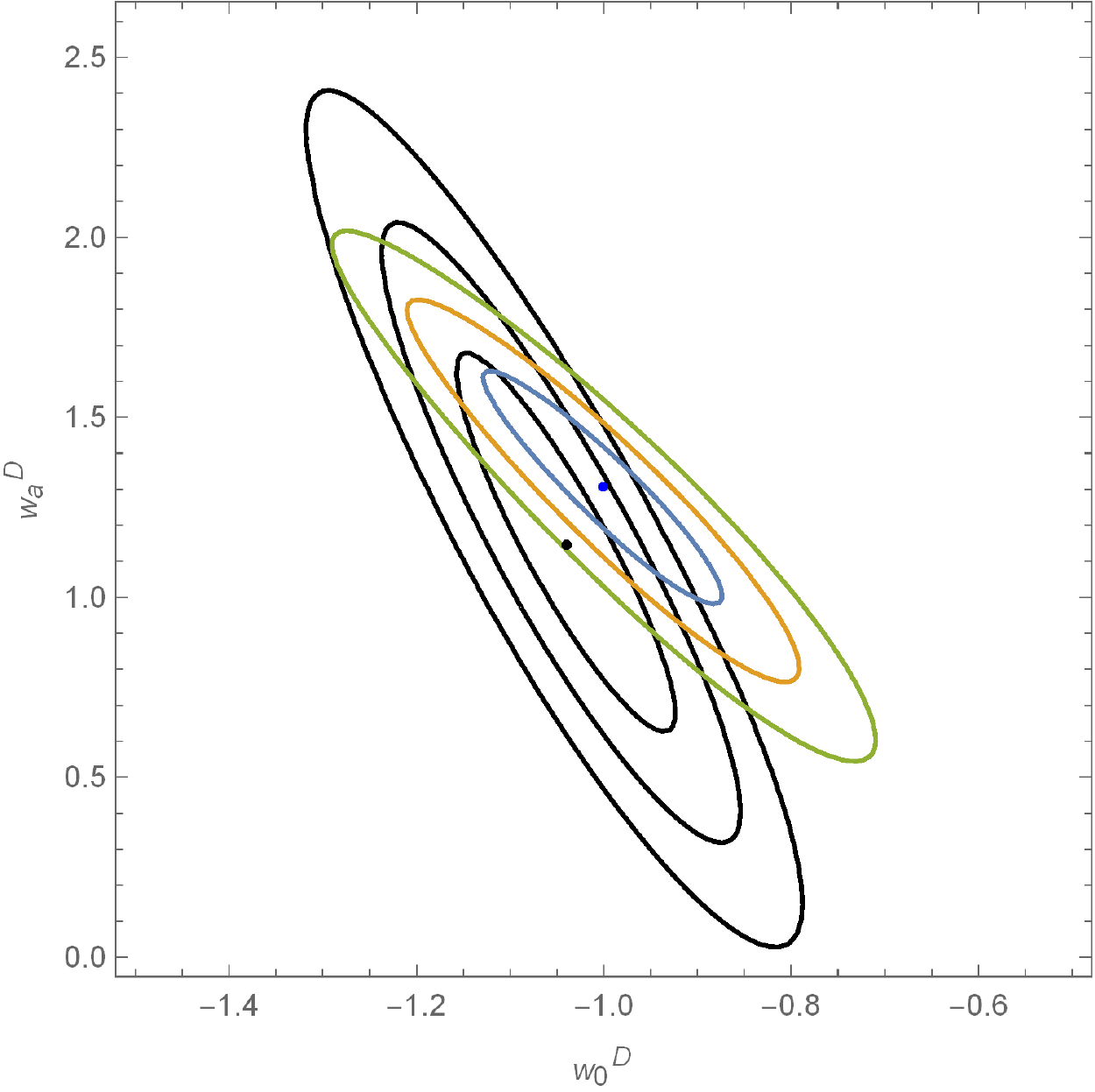}}
\end{minipage}
\caption{The 1$\sigma$, 2$\sigma$ and 3$\sigma$ confidence regions of the parameters $\Omega_{m}^{\now\CD}$, $w_0^{\CD}$
, and $w_a^{\CD}$ for the CPL backreaction model with the template metric using both SN Ia data and OHD. For the SN Ia data case, the contours are black, and the best fitting points are (0.32, $-$1.04), (0.32, 1.15) and ($-$1.04, 1.15) for the parameter pairs ($\Omega_{m}^{\now\CD}$, $w_0^{\CD}$), ($\Omega_{m}^{\now\CD}$, $w_a^{\CD}$) and ($w_0^{\CD}$, $w_a^{\CD}$) respectively. For the OHD case, the contours are in three different colours as one can see, and the best fitting points are (0.05, $-$1), (0.05, 1.31) and ($-$1, 1.31) for the parameter pairs ($\Omega_{m}^{\now\CD}$, $w_0^{\CD}$), ($\Omega_{m}^{\now\CD}$, $w_a^{\CD}$) and ($w_0^{\CD}$, $w_a^{\CD}$) respectively.}
\label{fig:1}
\end{figure}

\begin{figure}[h]
%\begin{tabular}{cc}
\begin{minipage}{0.45\linewidth}
  \centerline{\includegraphics[width=1\textwidth]{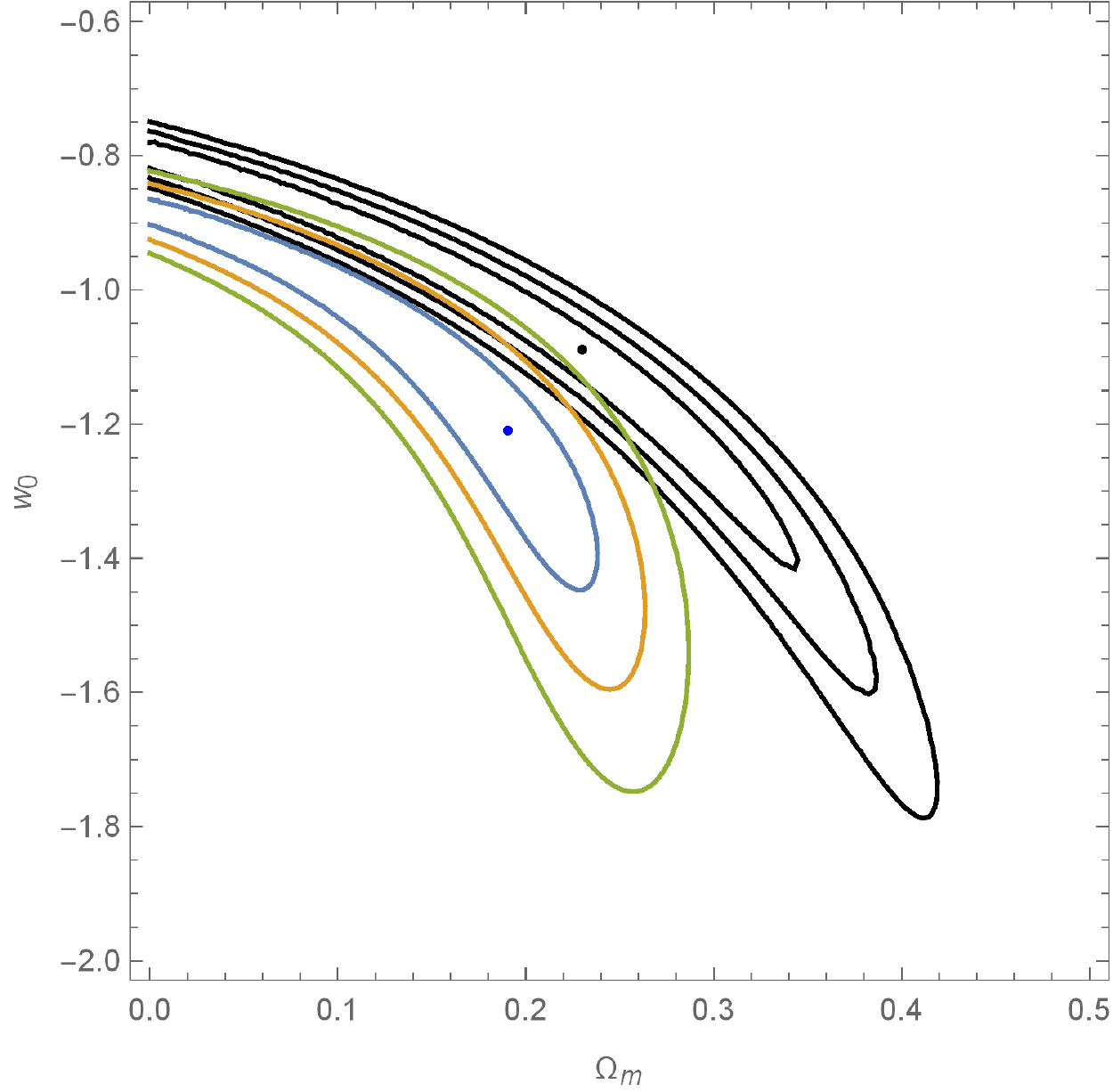}}
  \label{a}
\end{minipage}
\begin{minipage}{0.45\linewidth}
  \centerline{\includegraphics[width=1\textwidth]{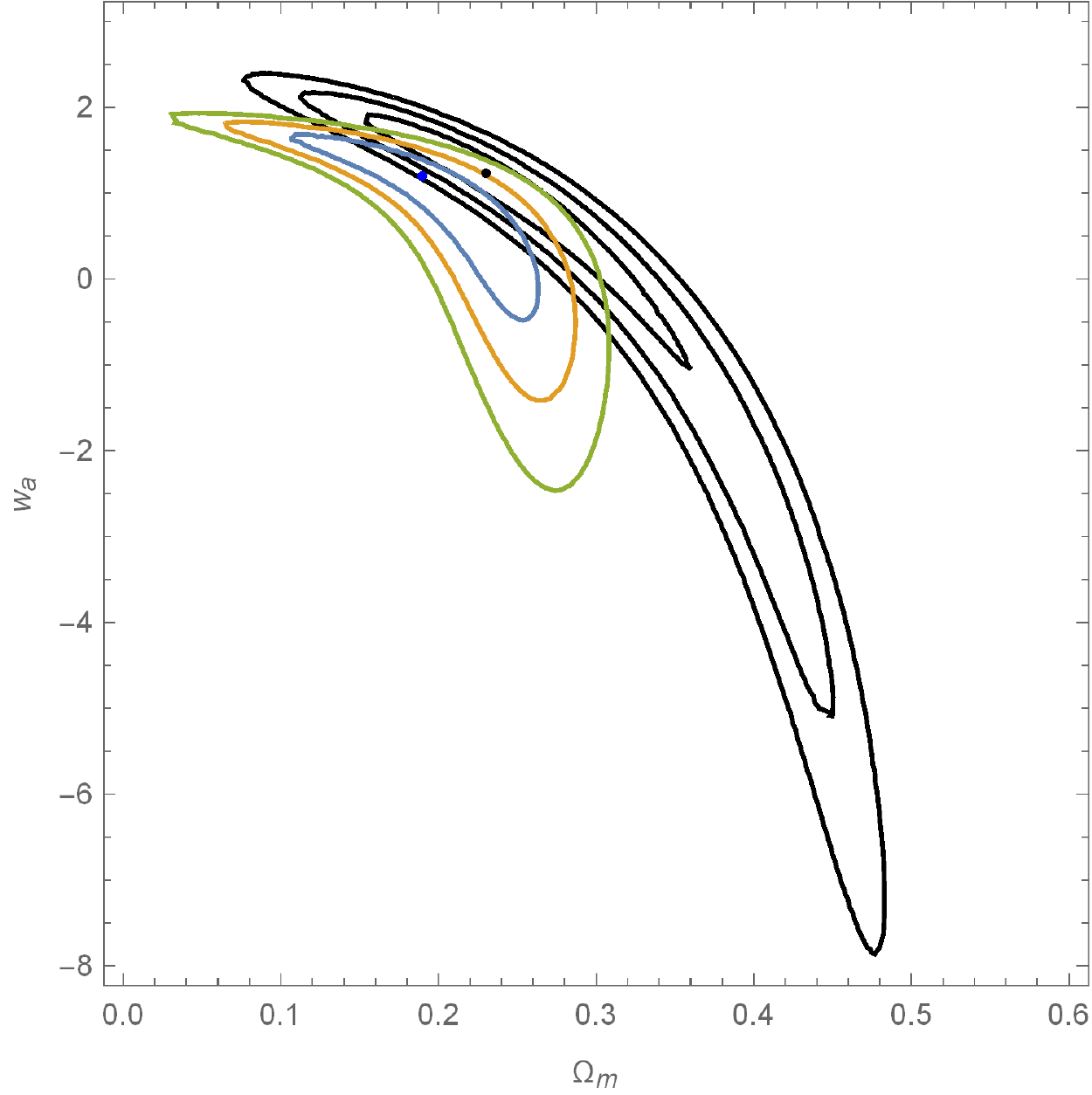}}
  \label{a}
\end{minipage}
\begin{minipage}{0.45\linewidth}
  \centerline{\includegraphics[width=1\textwidth]{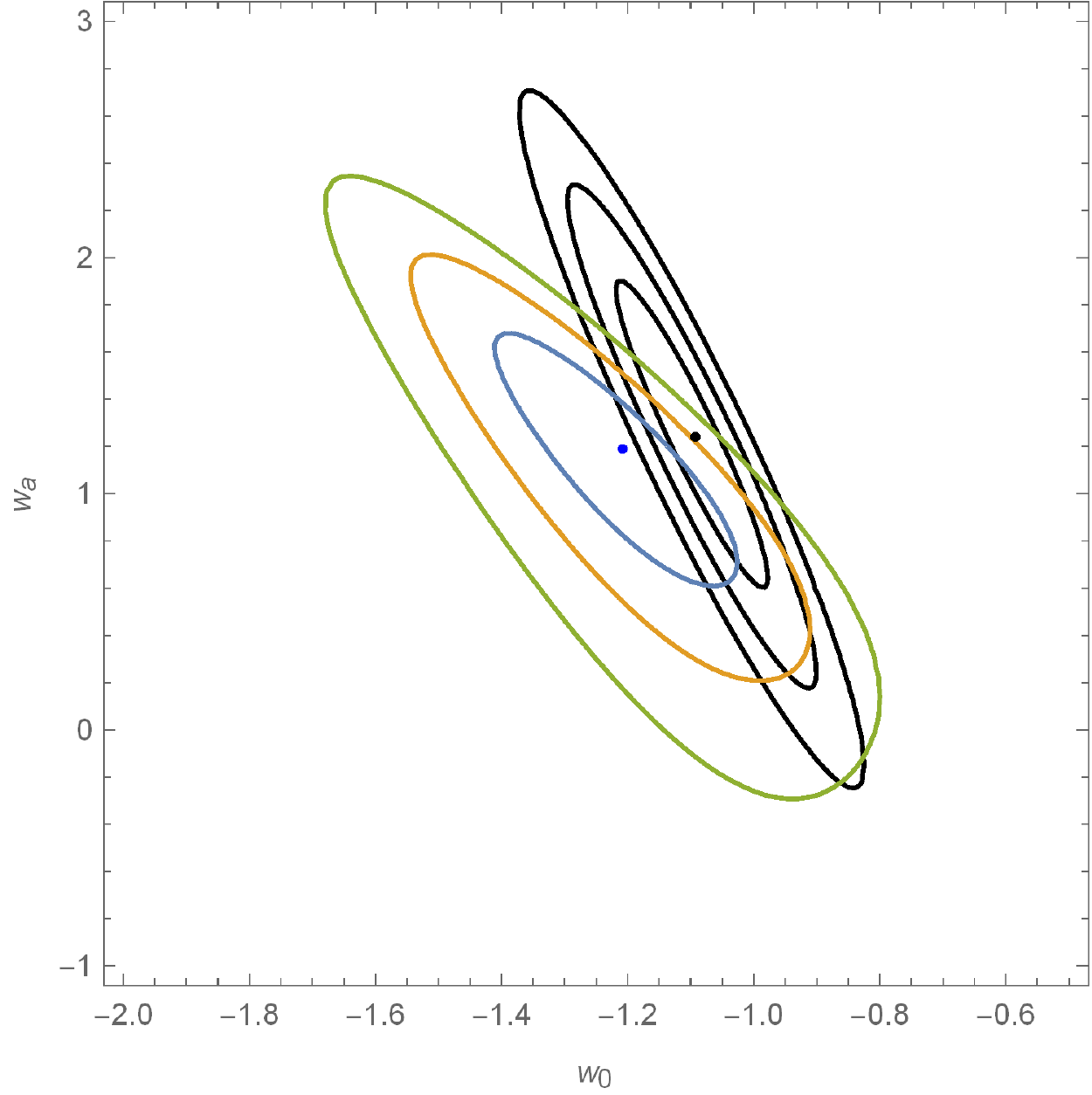}}
\end{minipage}
\caption{The 1$\sigma$, 2$\sigma$ and 3$\sigma$ confidence regions of the parameters $\Omega_{m}^{\now\CD}$, $w_0^{\CD}$
, and $w_a^{\CD}$ for the CPL backreaction model with the FLRW metric using both SN Ia data and OHD. For the SN Ia data case, the contours are black, and the best fitting points are (0.23, $-$1.09), (0.23, 1.24) and ($-$1.09, 1.24) for the parameter pairs ($\Omega_{m}^{\now\CD}$,
$w_0^{\CD}$), ($\Omega_{m}^{\now\CD}$, $w_a^{\CD}$) and ($w_0^{\CD}$, $w_a^{\CD}$) respectively. For the OHD case, the contours are in three different colours as one can see, and the best fitting points are (0.19, $-$1.21), (0.19, 1.19) and ($-$1.21, 1.19) for the parameter pairs ($\Omega_{m}^{\now\CD}$, $w_0^{\CD}$), ($\Omega_{m}^{\now\CD}$, $w_a^{\CD}$) and ($w_0^{\CD}$, $w_a^{\CD}$) respectively. }
\label{fig:2}
\end{figure}

Table \ref{2} show the constriant results of the CPL backreaction model with both the FLRW metric and template metric by using SN Ia data and OHD respectively, and Table \ref{3} show the prior of the parameters in the CPL backreaction model with both the FLRW metric and template metric. From the given fitting results one can infer that, for the template metric case, the tensions in parameters $\Omega_{m}^{\now\CD}$, $w_0^{\CD}$ and $w_a^{\CD}$ between two different data sets are 2.91$\sigma$, 0.05$\sigma$ and 0.49$\sigma$ respectively, and for the FLRW metric case, the corresponding tensions are 1.7$\sigma$, 0.17$\sigma$ and 0.07$\sigma$ respectively.
It is obvious that the CPL backreaction model with the FLRW metric has smaller parameter tensions in generally, and these tensions can actually be attributed to the lack of precision caused by the insufficient amount of OHD. The difference in the degree of parameter tensions in two different cases can also be inferred from Figure \ref{fig:1} and Figure \ref{fig:2}. In \cite{cao2018testing}, Cao et. al. show with the scaling backreaction model that, for the template metric case, their fitting results, which is produced by using OHD, is in contrary with the constraint results of Larena et al.\cite{larena2009testing}, which is produced by using SN Ia data and the position of CMB peaks, and for the FLRW metric case, two constraint results are basically consistent. What our work further show is that even for a more realistic backreaction model like the CPL backreaction model, similar results reproduce. Therefore, we draw the same conclusion that Cao et. al make that the prescription of $\kappa_{\CD}$ should be modified, or some other scenarios should be introduced.

\begin{spacing}{}
\begin{table}
\centering
\setlength{\tabcolsep}{0.5mm}{
\begin{tabular}{|lccc|}
\hline
{$z$}   & $H(z)$ & method & Ref.\\
\hline
$0.0708$   &  $69.0\pm19.68$      &  I    &  Zhang et al. (2014)-\cite{Zhang2014}   \\
    $0.09$       &  $69.0\pm12.0$        &  I    &  Jimenez et al. (2003)-\cite{Jimenez2003}   \\
    $0.12$       &  $68.6\pm26.2$        &  I    &  Zhang et al. (2014)-\cite{Zhang2014}  \\
    $0.17$       &  $83.0\pm8.0$          &  I    &  Simon et al. (2005)-\cite{Simon2005}     \\
    $0.179$     &  $75.0\pm4.0$          &  I    &  Moresco et al. (2012)-\cite{Moresco2012}     \\
    $0.199$     &  $75.0\pm5.0$          &  I    &  Moresco et al. (2012)-\cite{Moresco2012}     \\
    $0.20$         &  $72.9\pm29.6$        &  I    &  Zhang et al. (2014)-\cite{Zhang2014}   \\
    $0.240$     &  $79.69\pm2.65$      &  II   &  Gazta$\tilde{\rm{n}}$aga et al. (2009)-\cite{Gaztanaga2009}   \\
    $0.27$       &  $77.0\pm14.0$        &  I    &    Simon et al. (2005)-\cite{Simon2005}   \\
    $0.28$       &  $88.8\pm36.6$        &  I    &  Zhang et al. (2014)-\cite{Zhang2014}   \\
    $0.35$       &  $84.4\pm7.0$          &  II   &   Xu et al. (2013)-\cite{Xu2013}  \\
    $0.352$     &  $83.0\pm14.0$        &  I    &  Moresco et al. (2012)-\cite{Moresco2012}   \\
    $0.3802$     &  $83.0\pm13.5$        &  I    &  Moresco et al. (2016)-\cite{Moresco2016}   \\
    $0.4$         &  $95\pm17.0$           &  I    &  Simon et al. (2005)-\cite{Simon2005}     \\
    $0.4004$     &  $77.0\pm10.2$        &  I    &  Moresco et al. (2016)-\cite{Moresco2016}   \\
    $0.4247$     &  $87.1\pm11.2$        &  I    &  Moresco et al. (2016)-\cite{Moresco2016}   \\
    $0.43$     &  $86.45\pm3.68$        &  II   &  Gaztanaga et al. (2009)-\cite{Gaztanaga2009}   \\
    $0.44$       & $82.6\pm7.8$           &  II   &  Blake et al. (2012)-\cite{Blake2012}  \\
    $0.4497$     &  $92.8\pm12.9$        &  I    &  Moresco et al. (2016)-\cite{Moresco2016}   \\
    $0.4783$     &  $80.9\pm9.0$        &  I    &  Moresco et al. (2016)-\cite{Moresco2016}   \\
    $0.48$       &  $97.0\pm62.0$        &  I    &  Stern et al. (2010)-\cite{Stern2010}     \\
    $0.57$       &  $92.4\pm4.5$          &  II   &  Samushia et al. (2013)-\cite{Samushia2013}   \\
    $0.593$     &  $104.0\pm13.0$      &  I    &  Moresco et al. (2012)-\cite{Moresco2012}   \\
    $0.6$         &  $87.9\pm6.1$          &  II   &  Blake et al. (2012)-\cite{Blake2012}   \\
    $0.68$       &  $92.0\pm8.0$          &  I    &  Moresco et al. (2012)-\cite{Moresco2012}   \\
    $0.73$       &  $97.3\pm7.0$          &  II   &  Blake et al. (2012)-\cite{Blake2012}  \\
    $0.781$     &  $105.0\pm12.0$      &  I    &  Moresco et al. (2012)-\cite{Moresco2012}   \\
    $0.875$     &  $125.0\pm17.0$      &  I    &  Moresco et al. (2012)-\cite{Moresco2012}   \\
    $0.88$       &  $90.0\pm40.0$        &  I    &  Stern et al. (2010)-\cite{Stern2010}     \\
    $0.9$         &  $117.0\pm23.0$      &  I    &  Simon et al. (2005)-\cite{Simon2005}  \\
    $1.037$     &  $154.0\pm20.0$      &  I    &  Moresco et al. (2012)-\cite{Moresco2012}   \\
    $1.3$         &  $168.0\pm17.0$      &  I    &  Simon et al. (2005)-\cite{Simon2005}     \\
    $1.363$     &  $160.0\pm33.6$      &  I    &  Moresco (2015)-\cite{Moresco2015}  \\
    $1.43$       &  $177.0\pm18.0$      &  I    &  Simon et al. (2005)-\cite{Simon2005}     \\
    $1.53$       &  $140.0\pm14.0$      &  I    &  Simon et al. (2005)-\cite{Simon2005}     \\
    $1.75$       &  $202.0\pm40.0$      &  I    &  Simon et al. (2005)-\cite{Simon2005}     \\
    $1.965$     &  $186.5\pm50.4$      &  I    &   Moresco (2015)-\cite{Moresco2015}  \\
    $2.34$       &  $222.0\pm7.0$        &  II   &  Delubac et al. (2015)-\cite{Delubac2015}   \\
\hline
\end{tabular}}
\caption{The current available OHD dataset. The method I is the differential ages method, and II represents the radial Baryon acoustic oscillation (BAO) method. H(z) is in unit of ${\rm km/s/Mpc}$ here.
\label{tab:1} }
\end{table}
\end{spacing}{}

\begin{table}[!t]
\begin{minipage}[!t]{\columnwidth}
  \renewcommand{\arraystretch}{1.3}
  \centering
  \setlength{\tabcolsep}{0.6mm}{
  \begin{tabular}{|c|  c|c|}
\hline
                               &  SN Ia    & OHD
                               \\
  \hline
        \multicolumn{3}{|c|}{CPL backreaction model with template metric} \\
        \hline
$\Omega_{m}^{\now\CD}$         & $ 0.36_{-0.08}^{+0.05}$    &   $0.12 _{-0.04}^{+0.02}$
                     \\
$w_0^{\CD}$           &$-1.03_{-0.12}^{+0.14} $    & $-1.02_{-0.12}^{+0.10} $
    \\
$w_a^{\CD}$           &$1.51_{-0.48}^{+0.17} $    & $ 1.24_{-0.32}^{+0.26}$
    \\
 \hline $\chi^{2}_{min}$    &  $1034.033 $ &  $-8228.50 $
                      \\
 \hline
        \multicolumn{3}{|c|}{CPL backreaction model with FLRW metric} \\
        \hline
$\Omega_{m}^{\now\CD}$         & $ 0.38_{-0.08}^{+0.05}$    &   $ 0.24_{-0.02}^{+0.02}$
                     \\
$w_0^{\CD}$           &$-1.20_{-0.15}^{+0.18} $    & $ -1.16_{-0.16}^{+0.17} $
    \\
$w_a^{\CD}$           &$ 1.20_{-0.29}^{+0.27}$    & $ 1.24_{-0.52}^{+0.28} $
    \\
 \hline $\chi^{2}_{min}$    & $1032.19  $  &  $-8228.06 $
                      \\
  \hline
\end{tabular}
\caption{Fitting results of the CPL backreaction model with both the FLRW metric and template metric by using SN Ia data and OHD respectively.}
  \label{2}}
  \end{minipage}
\\[12pt]%设置两个表格之间的空白行距离
\begin{minipage}[!t]{\columnwidth}
  \renewcommand{\arraystretch}{1.3}
  \centering
  \setlength{\tabcolsep}{0.8mm}{
  \begin{tabular}{|c|  c|}
\hline
        \multicolumn{2}{|c|}{CPL backreaction model with template metric} \\
        \hline
$\Omega_{m}^{\now\CD}$         & $ [0.01,0.99]$
                     \\
$w_0^{\CD}$           &$[-2,2] $
    \\
$w_a^{\CD}$           &$[-3,3] $
    \\
 \hline
        \multicolumn{2}{|c|}{CPL backreaction model with FLRW metric} \\
        \hline
$\Omega_{m}^{\now\CD}$         & $  [0.01,0.99]$
                     \\
$w_0^{\CD}$           &$[-2,2] $
    \\
$w_a^{\CD}$           &$ [-10,3]$
    \\
  \hline
\end{tabular}
\caption{Prior of the parameters in the CPL backreaction model with both the FLRW metric and template metric.}
  \label{3}}
  \end{minipage}
%设置第二个表格与正文下文的空白行距离
\end{table}

\section{conclusion}
\label{sec:4}
In this paper, in order to further prove the conclusion drawn by Cao et. al. in \cite{cao2018testing} that the prescription of $\kappa_{\CD}$ needs to be modified, we equivalently regard the backreaction effects as the direct results of an effective perfect fluid whose EoS is parameterized as CPL form by us to model the realistic evolution of backreaction as accurately as possible. As a comparison, we employ two metrics to describe the late time Universe, one of them is the FLRW metric, and the other is the template metric that contain the geometrical instantaneous spatially-constant curvature $\kappa_{\CD}$ whose prescription needs to be tested. We then fit such CPL backreaction model using both SN Ia data and OHD with these two metrics, and from the the constraint results we find that, for the template metric case, the tensions in parameters $\Omega_{m}^{\now\CD}$, $w_0^{\CD}$ and $w_a^{\CD}$ between two different data sets are 2.91$\sigma$, 0.05$\sigma$ and 0.49$\sigma$ respectively, and for the FLRW metric case, the corresponding tensions are 1.7$\sigma$, 0.17$\sigma$ and 0.07$\sigma$ respectively. Obviously, the CPL backreaction model with the FLRW metric has smaller parameter tensions in generally, and these tensions can actually be attributed to the insufficient amount of OHD, however, for the other case, tensions are too large to be attributed to the insufficient amount of OHD. Therefore, we believe that the prescription of $\kappa_{\CD}$ needs to be modified.
\section*{Acknowledgments}
The paper is partially supported by the Natural Science Foundation of China.

\bibliographystyle{spphys}
\bibliography{a}

\end{document}